\documentclass[aps,twocolumn,floatfix, showkeys, superscriptaddress, nofootinbib]{revtex4}
\usepackage{amsmath}
\usepackage{graphicx,bm,hhline}
\usepackage{calrsfs}
\DeclareMathAlphabet{\pazocal}{OMS}{zplm}{m}{n}

\newcommand{\bcri}{{\bm R}_i}
\newcommand{\bk}{{\bm k}}

\newcommand{\br}{{\bm r}}
\newcommand{\brp}{{\br}^\prime}

\newcommand\sss{\scriptscriptstyle}

\newcommand{\wig}[1]{\mathrel{\hbox{\hbox to 0pt{\lower.6ex\hbox{$\sim$}\hss}\raise.4ex\hbox{$#1$}}}}

\begin{document}
\title{High Temperature Electronic Structure with KKR}

\author{C. E. Starrett}
\email{starrett@lanl.gov}
\affiliation{Los Alamos National Laboratory, P.O. Box 1663, Los Alamos, NM 87545, U.S.A.}

\date{\today}
\begin{abstract}
Modeling high temperature (10's or 100's of eV), dense plasmas is challenging due to the multitude of
non-negligible physical effects including significant partial ionization and multi-site
effects.  These effects cause the breakdown or intractability of common methods and approximations used
at low temperatures, such as pseudopotentials or plane wave basis sets.  Here we explore
the KKR Green's function method at these high temperature conditions.  The method is all-electron, does
not rely on pseudopotentials, and uses a spherical harmonic basis set, and so avoids
the aforementioned limitations.  It is found to be accurate for solid density aluminum and iron
plasmas compared when compared to a plane wave method at low temperature, while being able
to access high temperatures.
\end{abstract}
\pacs{ }
\keywords{warm dense matter, KKR, EOS}
\maketitle

\section{Introduction}
Material properties at high temperatures such as equation of state and opacity are 
used to model a diverse range of physical phenomena such inertial
fusion experiments \cite{haan11}, white dwarf stars \cite{fontaine01} and main sequence stars \cite{bailey15}.
Often density functional theory based average atom models are used to generate the
data due to their computational efficiency and reasonable physical fidelity \cite{wilson06, piron3, scaalp, starrett13, perrot87, ovechkin16}.  Other
higher fidelity methods are available \cite{militzer15, hamel12}, however they have limitations such
as extreme computational expense or in which properties can be calculated.

In this work we explore the possibility of using the KKR (Korringa-Kohn-Rostoker) \cite{korringa47, kohn54, ebert11, papanikolaou02}
Green's function method for equation of state at high temperatures.  KKR has been shown to be accurate for
total energy calculations at normal conditions \cite{zeller98} but has not yet, to our knowledge,
been explored in the literature for high temperature materials (i.e. 10$^4$ to 10$^6$ K).
This study follows the exploratory work of Wilson {\it et al} \cite{wilson11} who first demonstrated
the possibly of using this method for high temperature materials.

While average atom models have reasonable physical fidelity, one key piece of missing
physics is multiple scattering.  In average atom models the boundary condition on the
electron wavefunctions is that they match the free electron form at the ion sphere.
In practice this means that such models assume that scattering electrons are
asymptotically free, which is clearly not true in dense materials, where electrons 
go on to scatter multiple times.  This lack of multiple scattering has a significant
effect on the calculated electronic structure and hence the predicted material properties.

KKR does include multiple scattering, indeed it is sometimes called multiple scattering
theory \cite{ebert11, faulkner79}.  The basic idea is that space is partitioned into space-filling, non-overlapping
polyhedra.  Inside each polyhedron the `single-site' problem is solved, which amounts to
evaluating the Kohn-Sham Density Functional Theory (DFT) equation for the regular and irregular solutions and phase
shifts.  Then by taking into account the differing frames of reference of each single site
solution, the multisite Green's function is constructed \cite{ham61}.  From this Green's function all
the usual electronic properties such as EOS and opacity can be calculated \cite{ebert11}.

Here we show that an average atom model is a special case of KKR
where multiple scattering has been ignored and the polyhedra are approximated
as spheres.  We then focus on two exploratory example, fcc aluminum and bcc iron
at high temperatures.  Our use of crystal structures, instead of a disordered fluid 
structure is justified at this early stage of research in that it is still an improvement
over the average atom model, and is useful because it allows us to test a key
approximation, that of multiple scattering basis set convergence.
Also, such systems, with hot thermal electrons and nuclei at or near their lattice 
positions, are already of physical interest since they occur
in ultra fast heating experiments such as at the X-ray Free Electron Laser facility
at LCLS \cite{vinko12}.

A further simplification of the present implementation is that we use the Muffin Tin (MT)
approximation.  In essence this assumes that, for calculating the Green's function, the 
effective one electron potential is spherically symmetric inside each polyhedron.  This
is unnecessary in the context of of KKR \cite{zabloudil00}, but does simplify the implementation.  Again this is 
justified at this early stage of research because it allows us to assess the effect of
multiple scattering.  It is worth noting that even though the MT approximation is
used, the electron density is not spherically symmetric in the polyhedra
because the multiple scattering boundary condition breaks the symmetry.

The structure of this paper is as follows.  In section \ref{sec_pm} we outline
the physical model that is being modelled and look at basis set convergence
for aluminum and iron plasmas.
In section \ref{sec_res} we show results for equation of state (EOS) and
density of states (DOS) for these same aluminum and iron plasmas.  We compare
to a commonly used average atom model as well as to state of the art plane-wave
DFT calculations.
Finally, in section \ref{sec_con} we draw our conclusions.

\section{Physical model\label{sec_pm}}
\begin{figure}
\begin{center}
\includegraphics[scale=0.55]{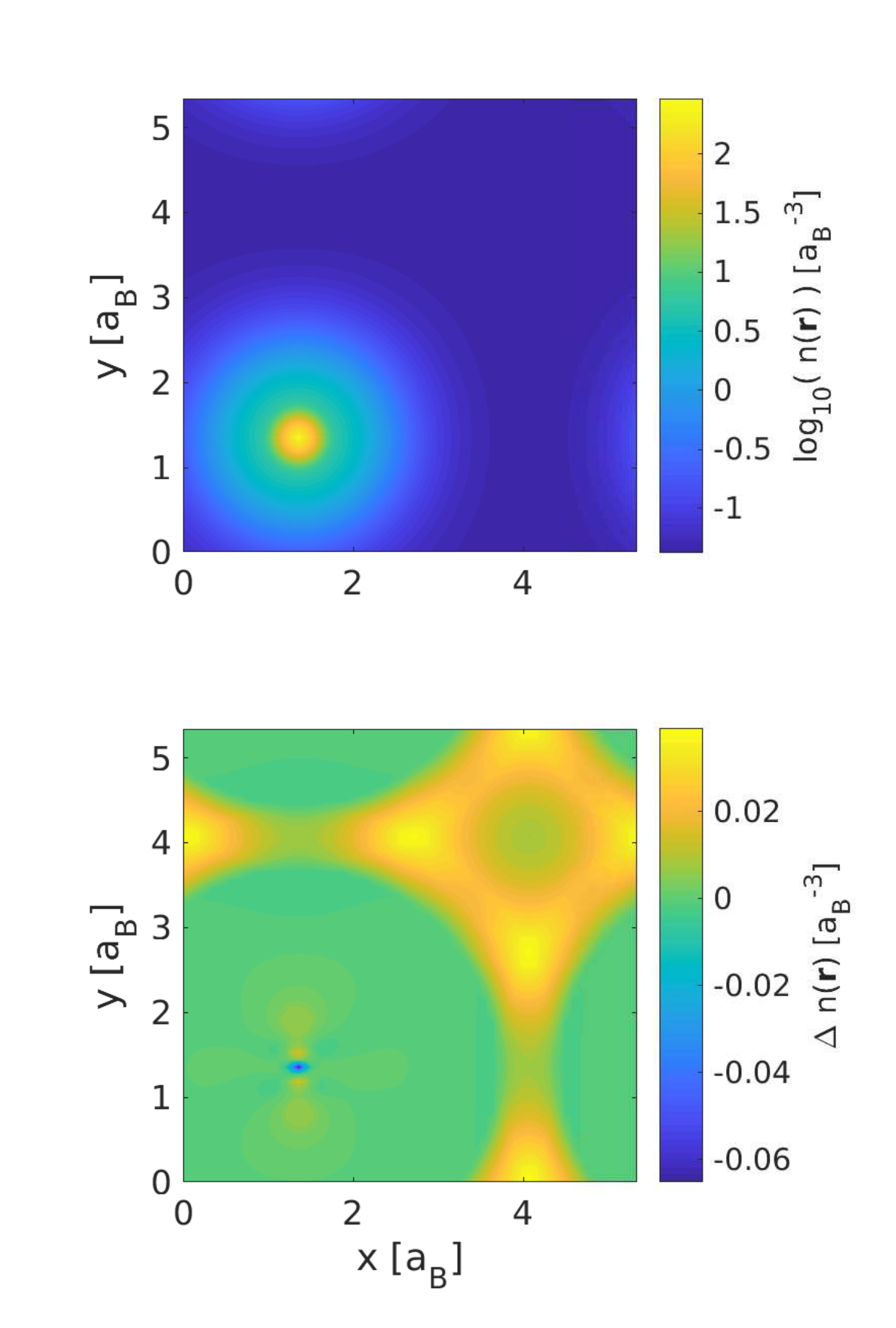}
\end{center}
\caption{(Color online) 
Top panel: a 2D slice of the 3D electron density $n_e(\br)$ for bcc iron at solid density and
a temperature of 10 eV.  Notice the logarithmic colorbar.  The electron density is
very strongly concentrated at the nuclear position.  Bottom panel: the difference $\Delta n_e$
between $n_e(\br)$ and the superposition density $n_e^{\sss super}(\br)$ that must be Fourier
transformed to solve the Poisson equation.
}
\label{fig_ne_iron}
\end{figure}
We consider dense plasmas in which thermalized electrons move in a background
of nuclei that are fixed at their lattice positions.  The
system is then periodic in the usual way.
The electron density $n_e(\br)$ is found using finite temperature density
functional theory \cite{mermin65}.  The Free energy functional is
\begin{equation}
F[n_e(\br)] = F^{\sss KS} + F^{\sss el} + F^{\sss xc}
\end{equation}
where $F^{\sss KS}$ is the kinetic-entropic term
\begin{equation}
F^{\sss ks} = U^{\sss k} - T S
\end{equation}
$U^{\sss k}$ is the electron kinetic energy contribution to the internal energy
\begin{equation}
U^{\sss k}  =  \int_{-\infty}^{\infty}d\epsilon f(\epsilon,\mu)\chi(\epsilon)\epsilon - 
            \int_{V}d^3r V^{eff}(\br) n_{e}({\br})
\label{uk}
\end{equation}
and $S$ is the entropy
\begin{eqnarray}
S & = & - k_B \int_{-\infty}^{\infty} d\epsilon 
\chi(\epsilon ) \nonumber\\
&&
\!\!\!\!\!\!\!\!\!\!\!\!\!\!\!\!\!\!\!\!
\!\!\!\!\!\!\!\!
\times \left[
f(\epsilon,\mu)\ln(f(\epsilon,\mu)) + (1-f(\epsilon,\mu))\ln(1-f(\epsilon,\mu)) 
\right]
\label{fs}
\end{eqnarray}
In these equations $\mu$ is the chemical potential found by requiring the
system volume $V$ to be charge neutral, $\epsilon$ is the energy, $f(\epsilon,\mu)$
is the Fermi-Dirac occupation factor, $\chi(\epsilon)$ is the density of states
per volume $V$ and $T$ is the temperature.

$F^{\sss el}$ is the electostatic free energy due to the Coulomb interactions of
the nuclei and electron density $n_e(\br)$ with each other and with themselves.  
$F^{\sss xc}$ is the exchange and correlation free energy for which we use
the temperature independent Perdew-Zunger form \cite{perdew81}.

Requiring $F$ to be at a minimum with respect to variations in $n_e(\br)$ subject to charge neutrality
leads to the effective one electron potential
\begin{equation}
V^{\sss eff}(\br) = V^{\sss el}(\br) + V^{\sss xc}(\br)
\end{equation}
where $V^{xc}(\br)$ is the exchange correlation potential and the electrostatic potential
is
\begin{equation}
V^{\sss el}(\br) = \sum_i \frac{-Z_i}{|\br -\bcri|} + \int d^3r\, \frac{n_e(\brp)}{|\br -\brp|}
\end{equation}
$Z_i$ is the nuclear charge of nucleus $i$.

This is a self-consistent field problem.  The solution procedure is as follows
\begin{enumerate}
\item
Input nuclear charges and positions, choose electron-exchange and correlation potential,
create initial guess at $n_e(\br)$.
\item
Solve Poisson equation for electrostatic potential $V^{el}(\br)$. 
\item
Calculate total effective potential $V^{\sss eff}(\br) $.
\item
Solve for an electron density in the presence of $V^{\sss eff}(\br)$.
\item
Mix input and output electron densities to get new input guess and repeat steps 2 to 5 until input and output
densities are the same to numerical tolerance.  
\end{enumerate}
An example of $n_e(\br)$ for bcc iron at solid density and a temperature of 10 eV is given in the top panel of
figure \ref{fig_ne_iron}.  The figure demonstrates the electron density is very strongly peaked near the nuclei,
due to the occupation of bound (or core) states.  To solve the Poisson equation a number of options are
available \cite{alam11, gonis91, vitos95, zabloudil_book}.  We have opted for the method presented in
reference \cite{starrett17} which uses fast-Fourier transforms (FFT) to solve the Poisson equation.  This method
is suitable for relatively small systems, with perhaps up to a few hundred unique nuclei only.  It works by splitting the
electron density into a sum of spherically symmetric electron densities placed at each nuclear site 
$n_e^{\sss super}(\br) = \sum_i n_{e,i}^{PA}(r)$
and a correction, $\Delta n_e(\br) = n_e(\br) - n_e^{\sss super}(\br)$.  The Poisson equation is then solved
by using a uniform 3D spatial grid to represent $\Delta n_e$, while the potential due to the
spherically symmetric quantities is solved for easily using a well known method \cite{starrett17}.
An example of the correction density is shown in the bottom panel of figure \ref{fig_ne_iron}.

To carry out spatial integrals like that in equation (\ref{uk}) we have used the method presented in \cite{alam11a}.

To generate a new guess at $n_e(\br)$ we use the 5$^{th}$ order extended Anderson's method due to Eyert \cite{eyert96, anderson65}.  This
is closely related to Broyden's method \cite{broyden65, johnson88} and typically converges in less than 10 iterations.

All that remains is a method to solve for the electron density.  This is where KKR is used.

\subsection{KKR for the electron density}
\begin{figure}
\begin{center}
\includegraphics[scale=0.55]{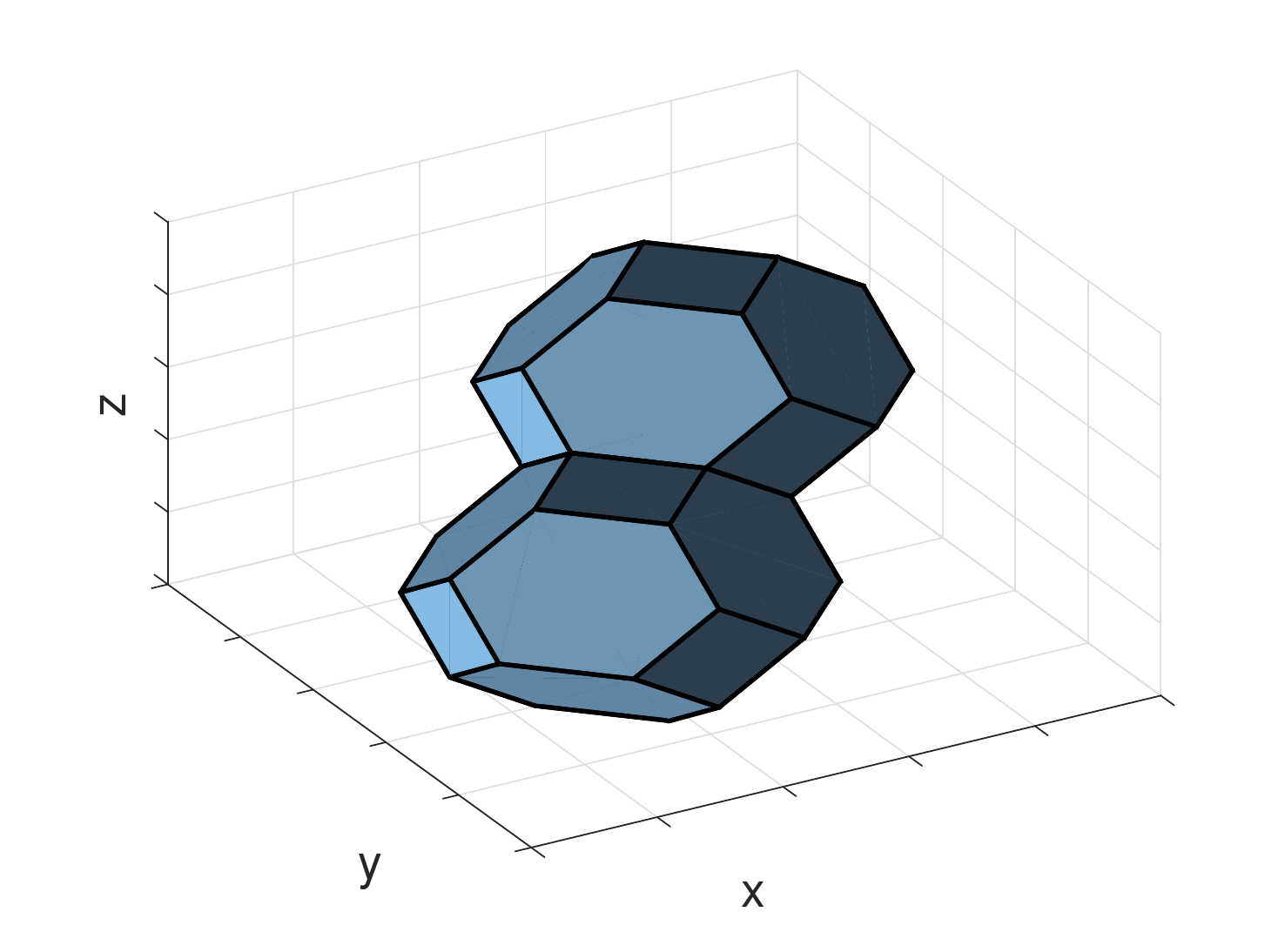}
\end{center}
\caption{(Color online) 
Voronoi decomposition of space around each nuclei here for a bcc crystal.  In KKR
the Schr\"odinger equation is solved inside each polyhedron.  These solutions together with 
the positions of the nuclei are then used to construct the Green's function for the entire system, from which 
the electron density and other material properties are calculated.
}
\label{fig_bcc}
\end{figure}

The electron density $n_e(\br)$ is expressed in terms of the Green's function  $G(\brp, \br, \epsilon)$
\begin{equation}
n_e(\br) = -\frac{1}{\pi}\Im \int_{-\infty}^{\infty}d\epsilon f(\epsilon,\mu)TrG(\br,\br,\epsilon)
\label{ne_gf}
\end{equation}
where the integral is over energy and non-relativistically the trace amounts to a factor of two that 
accounts for the spin degeneracy.  In KKR, the Green's function is constructed by
first partitioning space into non-overlapping polyhedra.  For materials with only one
type of nucleus we use a Voronoi decomposition, see figure \ref{fig_bcc}.
Inside each polyhedron the Schr\"odinger equation (or Kohn-Sham
equation in DFT) is solved using a spherical harmonic $Y_{lm}(\hat{\br})$ basis set, resulting in a set of regular $R_{l}(\epsilon, \br)$ and irregular 
$H_{l}(\epsilon, \br)$ wavefunctions \cite{starrett15}.  
These wavefunctions are normalized so that they match free electron behavior at the cell boundaries.  This also gives the 
T-matrix $\underline{t}(\epsilon)$ that is familiar from scattering physics, and is related to the scattering phase shifts.  
The T-matrix is a matrix in $l$ and $m$ and has size $(l_{max}+1)^2 \times (l_{max}+1)^2$, where $l_{max}$ is the maximum value
of $l$ included in the basis set expansion.  A super T-matrix is constructed for the $N$ particles in the super cell
$\underline{\bm t} = \{ \underline{t}^{nn^\prime} \delta_{nn^\prime}\}$ where $n=1,\ldots,N$ labels the nucleus \cite{zabloudil_book}.

\begin{figure}
\begin{center}
\includegraphics[scale=0.33]{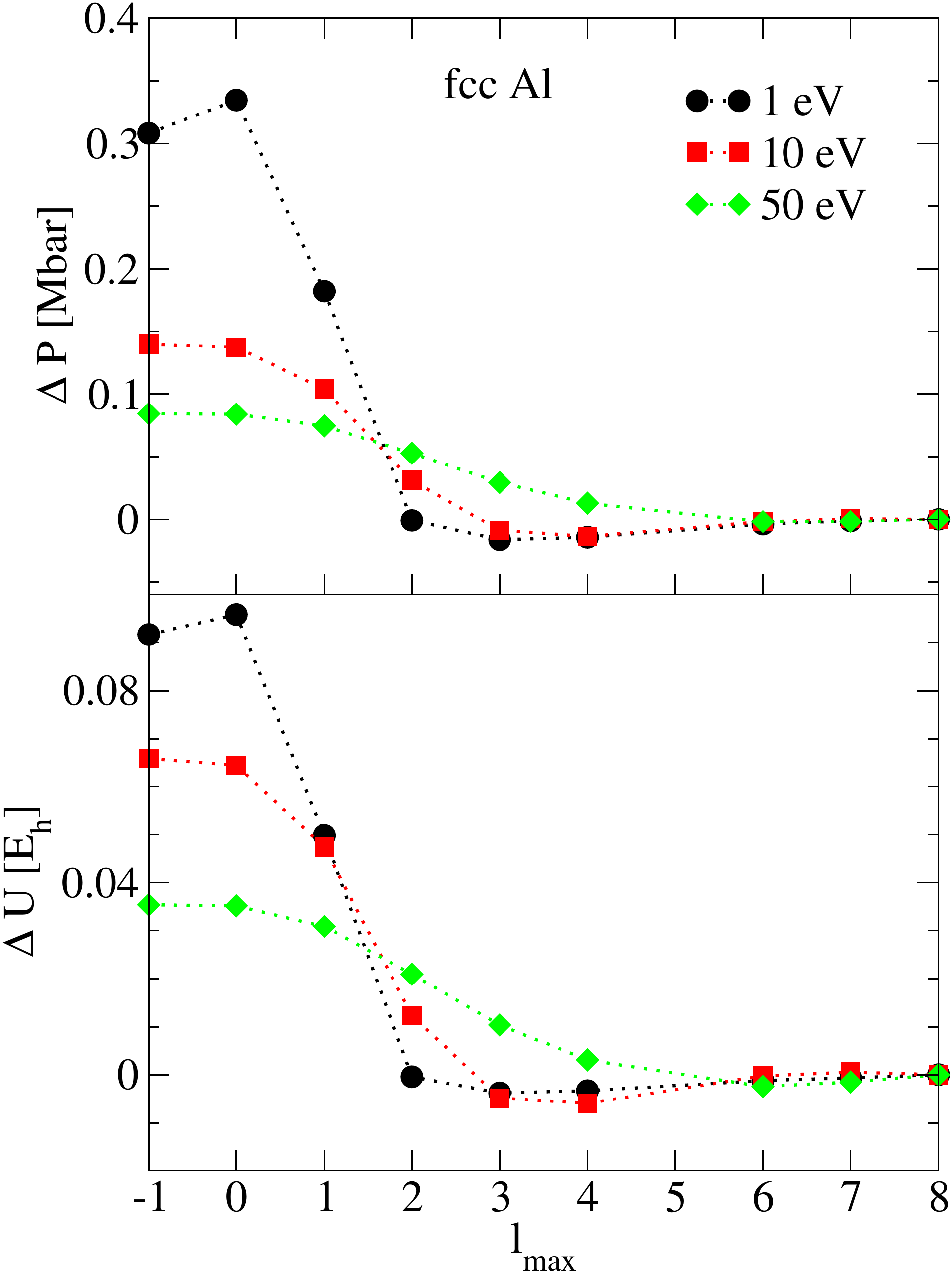}
\end{center}
\caption{(Color online) Convergence of changes in pressure $\Delta P$
and internal energy $\Delta U$ as a function of the truncation of the $l$ summation
for $G^{ms}$, for fcc aluminum.
}
\label{fig_lcon_al}
\end{figure}

The positions of the nuclei and the periodic nature of
the supercell are then encoded in the so-called structure constants matrix $\underline{G_0}(\epsilon,\bk)$ that accounts for the different 
origins of the basis sets.  $\underline{G_0}$ and $\underline{\bm t}$ are combined
in the structural Green's function matrix $\underline{\cal{G}}(\epsilon )$.  The trace of the Green's function
is then written 
\begin{equation}
Tr\,G(\br,\br,\epsilon) = G^{ss}(\br,\br,\epsilon) + G^{ms}(\br,\br,\epsilon)
\label{gf_sep}
\end{equation}
where the single-site part is
\begin{equation}
G^{ss}(\br,\br,\epsilon) = -2m \imath \sqrt{2m\epsilon} \sum_{l=0}^{l_{max}} \frac{2(2l+1)}{4\pi} R_l(\epsilon,\br) H_l(\epsilon,\brp)
\end{equation}
and the multi-site part is
\begin{equation}
\begin{split}
G^{ms}(\br,\br,\epsilon) = &2m 
\sum_{l=0}^{l_{max}} 
\sum_{l^\prime=0}^{l_{max}} 
\sum_{m=-l}^{l} 
\sum_{m^\prime=-l^\prime}^{l^\prime} 
{\cal G}^{nn}_{lm l^\prime m^\prime}
(\epsilon) \\
& \times R_l(\epsilon,\br) R_{l^\prime}(\epsilon,\br) Y_{lm}(\hat{\br}) Y_{l^\prime m^\prime}^*(\hat{\br})
\end{split}
\end{equation}
where we have limited ourselves to the local (i.e. $r=r^\prime$) expression
and we have assumed that the scattering potential inside each polyhedron is spherically
symmetric.  This assumption is an approximation that simplifies the numerics but is
not formally, or even practically, necessary \cite{zabloudil00}.  To enforce the spherical symmetry we
use the common Muffin Tin (MT) approximation.  In practice this is done by defining a 
MT sphere for each polyhedron, which is the largest sphere that completely fits into
the polyhedron.  Inside the spheres the potential $V^{eff}(\br)$ is spherically
averaged.  Outside the spheres, in the so called interstitial
region, $V^{eff}(\br)$ is replaced with its average value $\bar{V}^{\sss MT}$.  The total
MT potential is then shifted by $\bar{V}^{\sss MT}$ \cite{zabloudil_book}.  The
Schr\"odinger equation is then solved for this potential.
\begin{figure}[!]
\begin{center}
\includegraphics[scale=0.33]{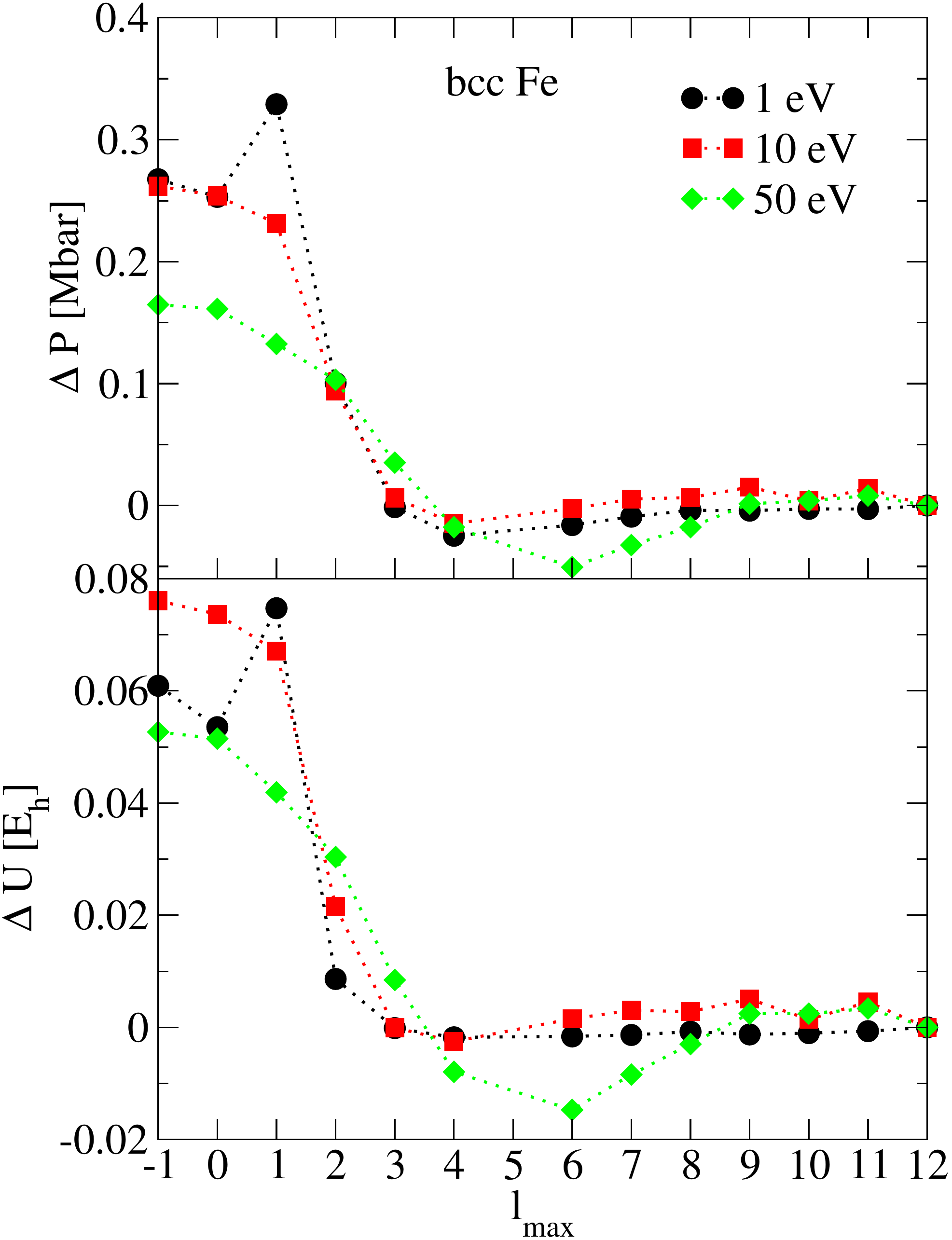}
\end{center}
\caption{(Color online) Convergence of changes in pressure $\Delta P$
and internal energy $\Delta U$ as a function of the truncation of the $l$ summation
for $G^{ms}$, for bcc iron.
}
\label{fig_lcon_fe}
\end{figure}

This is the only place the MT potential is used, i.e. to generate the Green's function.  It
is not (directly) used for example, when calculating the equation of state which 
explicitly depends on the potential $V^{eff}(\br)$.

It is interesting to note that the single-site and multi-site contributions to the
Green's function are fully separable, equation (\ref{gf_sep}).  Making the approximation that the multi-site
contributions to the Green's function and Poisson equation can be neglected, and approximating the polyhedra as spheres with volume
equal to the average volume per atom, we are left with an average atom model 
\cite{wilson06, liberman, starrett15}.  Such average atom models are widely used for
equation of state tables.  This derivation brings insight into the approximations
are missing physics inherent in average atom models. 

\subsection{Basis Set Truncation}
A key question of the method is that of the choice of $l_{max}$.  Since the
matrix that has to be inverted has size $N (l_{max}+1)^2$ \cite{thiess12} it is clearly
in our interest to keep $l_{max}$ as small as possible.  
To achieve this we treat $l_{max}$ for $G^{ss}$ and $G^{ms}$ separately.  $l_{max}$
for $G^{ss}$ can be numerically converged automatically by the code using a
common trick employed in average atom models \cite{blenski95,gill17}.
$l_{max}$ for $G^{ms}$ determines the size of the matrix to be inverted
and can be converged separately through convergence testing.
One can expect $l_{max}$ for $G^{ms}$  to converge at much smaller values than
that for $G^{ss}$ which for high temperature cases can need $l_{max} > 100$.
The reason for this is that electrons in states characterized by
large values of $l$ also have higher energies, and therefore behave more
like free electrons.  Since the t-matrix elements for free electrons
are zero, so is $G^{ms}$.
\begin{figure}
\begin{center}
\includegraphics[scale=0.33]{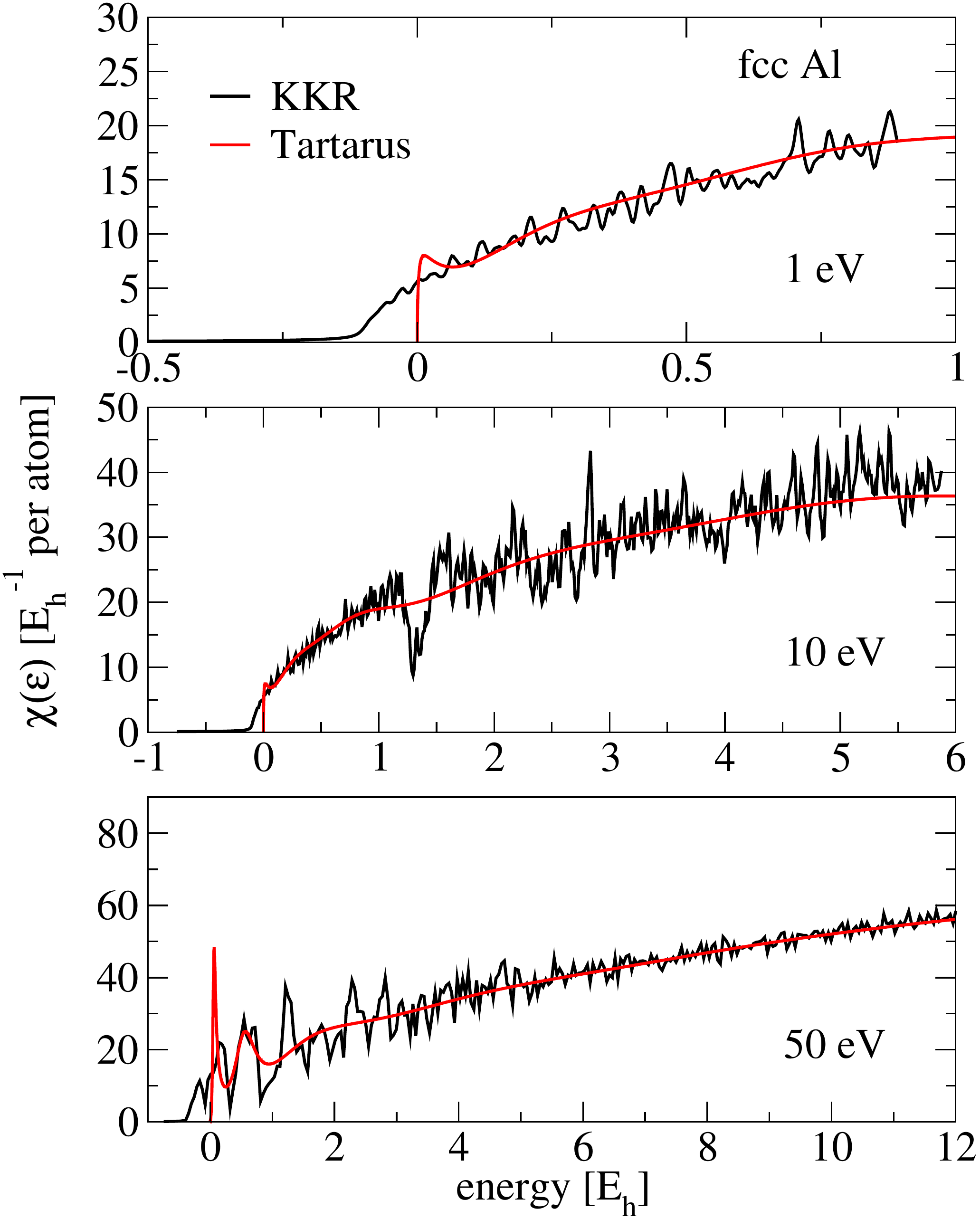}
\end{center}
\caption{(Color online) Density of states for fcc aluminum at 2.7 g/cm$^3$.  
}
\label{fig_dos_al}
\end{figure}

To test this expectation figures \ref{fig_lcon_al}  and \ref{fig_lcon_fe}
show convergence of the EOS for fcc aluminum and bcc iron respectively.  $l_{max}$ in
the figures is that used for $G^{ms}$ only, that for $G^{ss}$ is converged automatically by
the code.  $l_{max} = -1$ is used to represent the case where $G^{ms} = 0$.  For
both cases we consider electrons with temperatures of 1, 10 and 50 eV.  For all cases
the pressure and internal energies are well converged by $l_{max} = 8$.
Note that we have plotted absolute changes in pressure and energy; relative
changes would show a relatively small influence of multiple scattering on EOS as
temperature is increased (also see figures \ref{fig_eos_al} and \ref{fig_eos_fe}).

Another interesting feature of these plots is that for iron at 1 eV there is a `spike' in the EOS changes
for $l_{max}=2$, and rapid convergence thereafter.  This is due to capturing the
3d valence band feature in the multiple scattering treatment.  This 3d feature is
expected to be particularity important at low temperatures, and less so at high temperatures
(see later, figure \ref{fig_dos_fe}).

\section{Equation of State and Density of States\label{sec_res}}
\begin{figure}
\begin{center}
\includegraphics[scale=0.33]{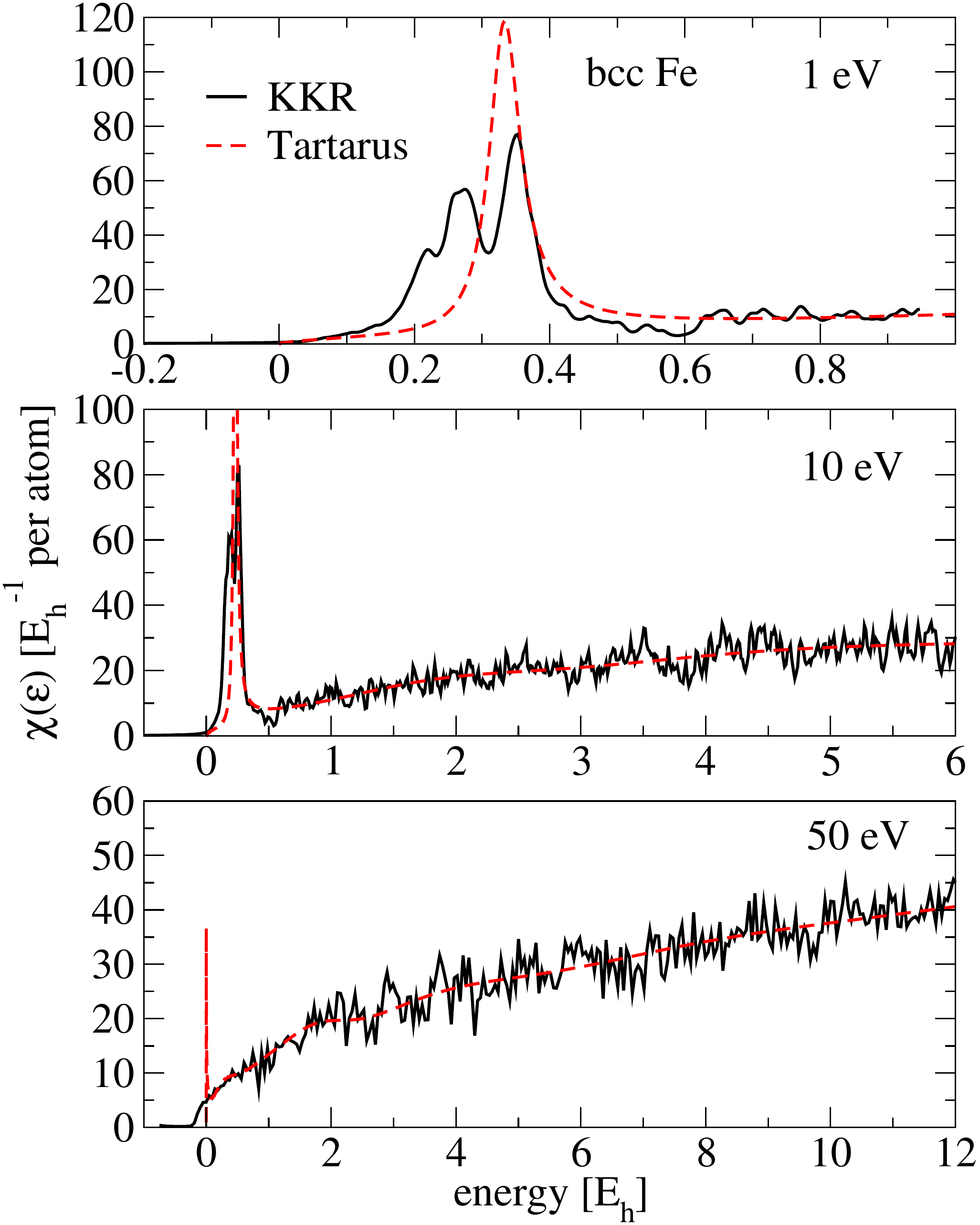}
\end{center}
\caption{(Color online) Density of states for bcc iron at 7.894 g/cm$^3$.  
}
\label{fig_dos_fe}
\end{figure}
In figures \ref{fig_dos_al} and \ref{fig_dos_fe} we plot density of states for fcc aluminum
and bcc iron respectively.  Compared are the DOS from KKR and that from the average atom code
{\texttt Tartarus} \cite{gill17}.  Firstly, we note that `noisy' KKR curves are due to the
fact that for real energies the KKR DOS is a series of Dirac delta functions at the eigenvalues
of the system.  What we actually plot is the real part of the DOS parallel to the real energy axis, with an
imaginary part of the energy equal to $0.01$ E$_h$.  This effectively broadens the DOS by
convolving it with a Lorentzian with a width equal to this imaginary part of the energy \cite{johnson84}.  Such
noise can also be caused by insufficient k-point sampling.  We have converged the shown DOS
so that any larger scale features are actual predictions of the model.
\begin{figure}
\begin{center}
\includegraphics[scale=0.33]{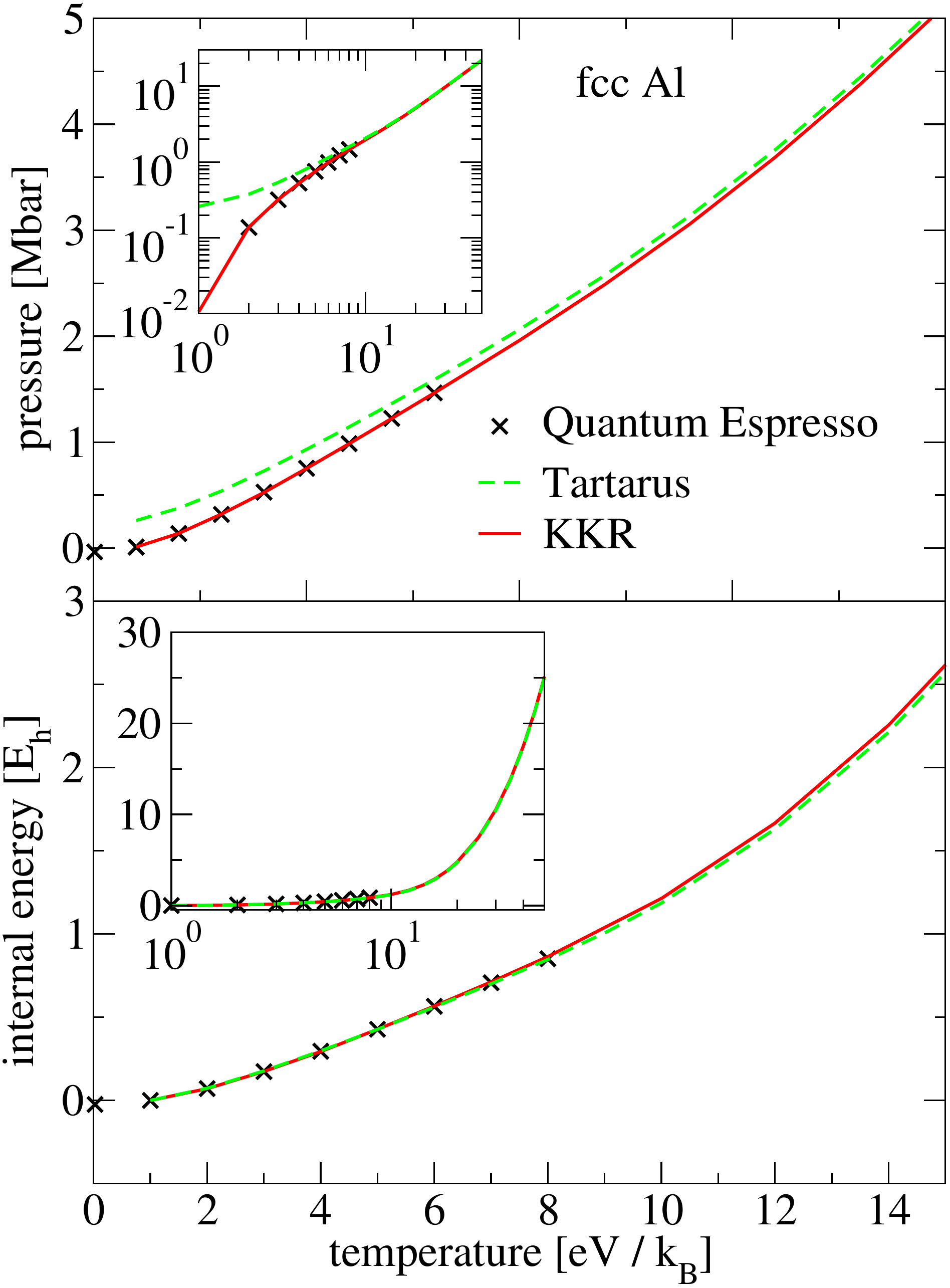}
\end{center}
\caption{(Color online) Equation of state of fcc aluminum at solid density.
}
\label{fig_eos_al}
\end{figure}

Generally speaking, there is a very good level of agreement between the KRR and average atom 
results.  For aluminum at 1 eV the most significant difference is near zero energy where
the KKR DOS extends to roughly -0.1 E$_h$, while the average atom goes to zero at zero energy,
as it must.  In the average atom model the only states that can appear at negative energies
are discrete bound states, whose DOS are Dirac delta functions at the eigenenergy.  Clearly KKR
does not have this restriction and the DOS looks much more free electron like.  Such differences
will show up in spectroscopic quantities like opacity.  For aluminum at 10 eV
one significant difference is a dip in the KKR DOS at 1 to 1.5 E$_h$ in contrast to a
very weak dip in the average atom result.  
At 50 eV for aluminum the KKR DOS displays structure at small energies that is 
quite different to that displayed by the average atom.  Clearly the more realistic
treatment of ionic structure and non-spherical symmetry in KKR are the cause of these
differences.  

For iron, figure \ref{fig_dos_fe}, the differences are even more pronounced.  At 1 eV
the large 3d resonance feature in the average atom curve is broadened and reduced in
height in KKR.  For 10 eV a similar result is seen.  By 50 eV the 3d state has recombined
and is no longer visible on this scale.  The average atom DOS has a large spike at small
positive energy associated with a nearly recombined s-state. This is completely
missing from the KKR DOS which is very free electron like.

In summary, there is a good level of general agreement between KKR and average atom but
significant differences in details caused by the more realistic treatment of ion structure
and non-spherical symmetry in the KKR model.

\begin{figure}
\begin{center}
\includegraphics[scale=0.33]{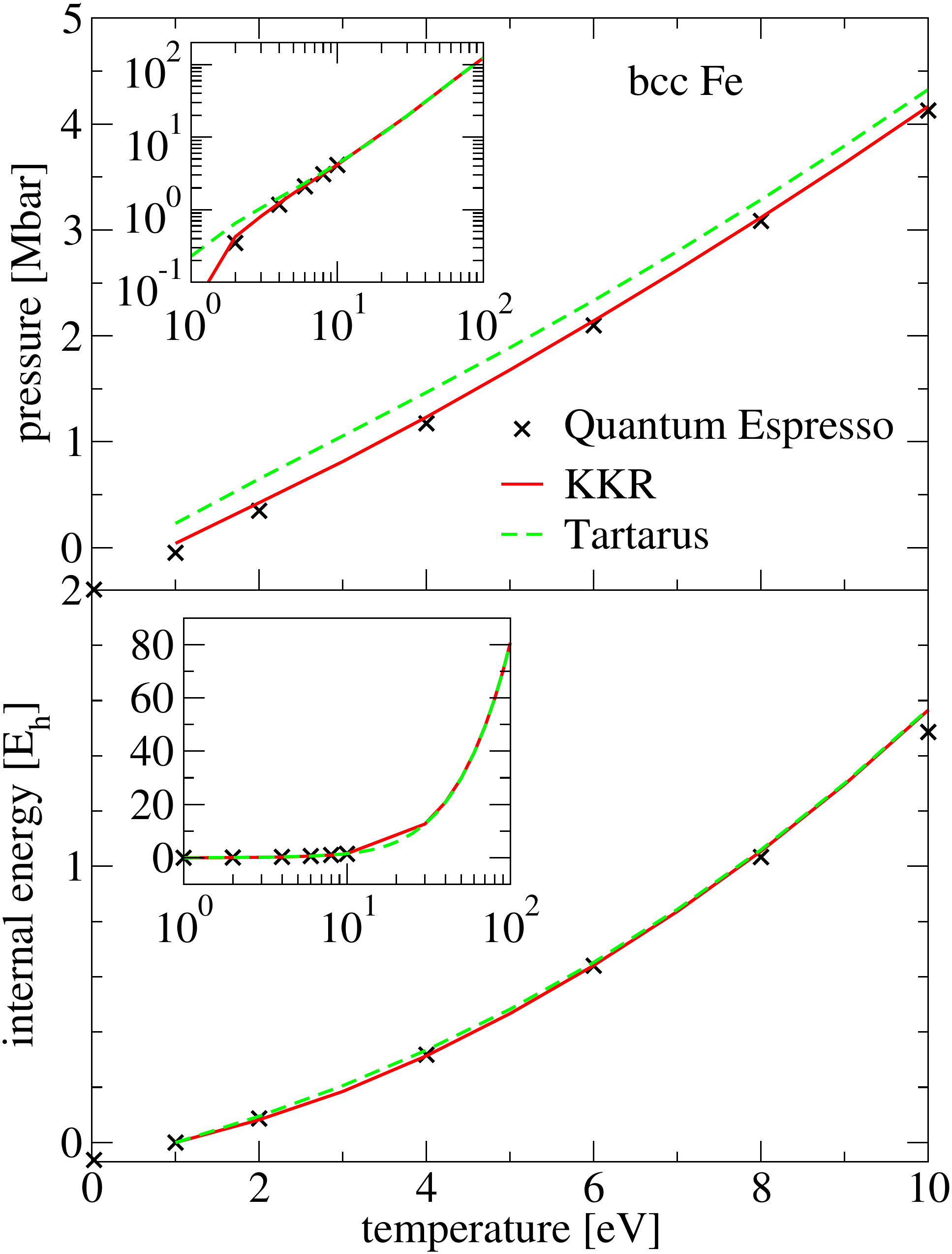}
\end{center}
\caption{(Color online) Equation of state of bcc iron at solid density.
}
\label{fig_eos_fe}
\end{figure}

In figures \ref{fig_eos_al} and  \ref{fig_eos_fe} EOS from the KKR model is compared
to both average atom results and, at low temperatures, to the plane-wave DFT code
Quantum Espresso \cite{qe2009, qe2017} (QE).  The QE calculations use pseudopotentials but are thought to be
accurate  at relatively low temperatures (up to 10 eV here).  Moreover, such plane wave type calculations
become prohibitively expensive as temperature increases.  KKR on the other hand can easily
access high temperatures (up to 100 eV here) without significant scaling.  Though the calculation
of $G^{ss}$ does scale with temperature, the prefactor is small and is not significant for any 
of the results presented here.  Due to the cutoff of $l_{max}$ for $G^{ms}$, its computational cost
does not increase with temperature, hence the quasi-temperature independence of the overall
computational cost of the method.

In figure \ref{fig_eos_al} we see very good agreement between the QE and KKR results for both pressure
and internal energy, and for higher temperatures, very good agreement between the average atom results
and KKR.  This is a key result: KKR is accurate at both low and high temperature
while remaining computationally feasible.  In figure \ref{fig_eos_fe}, for iron, similar trends are observed.
Now however, there are some relatively small differences between the KKR and QE results, presumable
due to the muffin-tin approximation that we have used.  Nevertheless, there is a clear
and substantial improvement over average atom, particularly for pressure.

\section{Conclusion \label{sec_con}}
The KKR method has been explored for use in dense plasmas.  It
was found that the method is accurate for equation of state when compared
to other state of the art methods at low temperature, while being able
to access high temperature, partially degenerate, plasma states without
prohibitive computational cost.  As such, it offers a promising capability
to provide high physical fidelity modelling of warm and hot dense plasmas.

It was also shown that the commonly used average atom model is a special
case of the KKR method, where multiple scattering has been ignored and
the polyhedra around each nucleus approximated by spheres.

Our relatively crude implementation of the method does not exploit many of
the numerous methodological and numerical improvements that have been presented
over its long history in solid state physics.  For example, we have used the 
obsolete muffin tin approximation due do its relative simplicity to implement.
This approximation is unnecessary \cite{zabloudil00}, but is adequate for
present purposes.  Also, we have used a non-relativistic implementation,
but relativity in the form of the Dirac equation is possible \cite{zabloudil_book}.
Also, the method can be made to scale linearly with number of particles \cite{zeller11_book}.
In summary, this method offer much promise to improve our understanding of
warm and hot dense plasmas.

\section*{Acknowledgments}
The author thanks and T. Sjostrom for doing the Quantum Espresso calculations for this work.
This work was performed under the auspices of the United States Department of Energy under contract DE-AC52-06NA25396.

\bibliographystyle{unsrt}
\bibliography{phys_bib}

\end{document}